\documentclass[conference]{IEEEtran}
\usepackage[ruled,vlined]{algorithm2e}
\usepackage{amsmath}
\usepackage{amssymb}
\usepackage{array}
\usepackage[justification=centering]{caption}
\usepackage{cite}
\usepackage{enumerate}
\usepackage{epsfig}
\usepackage{etex}
\usepackage{flafter}
\usepackage{graphics}
\usepackage{graphicx}
\usepackage{epstopdf}
\usepackage{subfig}
\usepackage{float}
\usepackage{booktabs}
\usepackage{soul}
\usepackage{mathtools}
\usepackage{array}
\usepackage{multicol}
\usepackage[normalem]{ulem}
\usepackage[english]{babel}
\usepackage{physics}
\usepackage{lipsum}
\usepackage{cleveref}
\usepackage[autostyle]{csquotes}

\usepackage{graphicx}
\begin{document}

\title{Silicon Photonics Testing: Design for Testability, Fault Detection, and Manufacturing Variation Analysis in Photonic Integrated Circuits}
\author{
	\IEEEauthorblockN{Pratishtha Agnihotri, Priyank Kalla, Steve Blair}
	\IEEEauthorblockA{Electrical \& Computer Engineering\\
          University of Utah, Salt Lake City, UT, USA\\
}}

\maketitle
\pagestyle{plain}
\thispagestyle{plain}
\begin{abstract}
	This paper proposes a design-for-test (DFT) methodology and
architecture for testing and validation of silicon photonic integrated
circuits. We describe the design of silicon photonic circuits and
components that comprise the proposed DFT architecture. The designs are
extensively simulated and validated as test-access and fault-detection
circuitry. We demonstrate how the DFT approach can be deployed on
photonic integrated circuits and how they can be tested for correct
operation, in terms of signal power and phase. The application is
demonstrated on two distinct types of designs -- an optical neural
network comprising optical devices in a feed-forward topology, and on
a optical logic circuit with feedback loops.  

\end{abstract}
\begin{IEEEkeywords}
Silicon photonics, design-for-test, Mach-Zehnder interferometer,
directional coupler, phase modulation, test access point
\end{IEEEkeywords}
\section{Introduction}
The field of silicon photonics has seen a marked increase in both
research endeavors and commercial ventures. This is largely driven by
notable improvements in the efficacy of photonic elements, and the
increase in scale and complexity of photonic integration over the last
decade. Silicon photonics is 
particularly well-suited for applications necessitating
high-bandwidth, cost-effective, and long-range interconnect
capabilities. This trend is further underscored by the exponential
expansion of data within the digital economy, driving innovations
across computing, storage, and networking technologies \cite{bowers},
\cite{HPC_photonics}, \cite{HPC_photonics1}. \\ 
\indent The adoption of silicon for photonics is underpinned by
several pivotal factors, with a notable emphasis on the high
refractive index contrast between silicon and silicon oxide, which
allows strong light confinement and a compact waveguide footprint in
the silicon-on-insulator (SOI) layer. Moreover, the availability of
high-speed modulation effected by carrier injection or extraction
makes silicon photonics appealing for embedding switching elements in
the routing/communication fabrics, thus bringing a convergence of
computation and communication -- in applications such as optical
neural networks \cite{Hochberg:ONN17}, optical logic
\cite{condrat2011logic}, etc.  These advantageous
characteristics are further benefited by 
established manufacturing processes inherent to the
CMOS ecosystem, which can be easily re-tooled and employed for silicon
PIC fabrication \cite{wim_umar}, \cite{GrahamTReed}.\\ 
\indent Along with the advantages, Si-photonics also brings about some
challenges. The significant material contrast inherent in silicon
photonic circuits, however, renders them highly susceptible to even
minor discrepancies in fabricated geometries, notably in the
dimensions such as width or thickness of submicron silicon
waveguides. This vulnerability becomes increasingly problematic with
the increasing circuit complexity, as variations propagate and build
up throughout the circuit. Such deviations affect both power and phase
of the optical signals, cause faulty interference patterns, and
inevitably result in performance deterioration and poor yield
\cite{pratishtha1},\cite{pratishtha12}, \cite{mahdi1},
\cite{variation_wim}.

There is a need for test automation in the
silicon photonics ecosystem to effectively test wafers, chips and
components in the optoelectronic domain. A dedicated {\it test
  circuit} embedded on the chip is imperative for real-time data
collection and analysis, detection of failure, and also crucial for
post-fabrication tuning/calibration of photonic integrated circuits
(PICs). Such a {\it design-for-test} (DFT) circuitry, along with
associated automatic test pattern generation (ATPG) techniques,  may
yield numerous advantages, including but not limited to time and cost
reduction in testing, enhanced testing coverage and reliability, as
well as improved scalability and yield for silicon PICs. 

{\it Contributions:} In this paper, we present a design-for-test (DFT)
architecture that can be employed on a silicon PIC to simplify its
testing for proper operation. In support of our DFT architecture, we
describe the design of silicon photonic circuits and components that
act as test-access and fault-detection circuitry. We show how these
circuits can be DFT-inserted at test access points, and
show how they can be employed for fault/failure detection. We describe
their design and implementation, and validate their proper operation
using extensive simulations and experiments. 

The manipulation of optical signals in PICs is accomplished by
employing on-chip optical components like waveguides, 
couplers, electro-optic modulators, photodetectors, lasers,
etc. \cite{programmable_photonics}. To deal with optical signals
on-chip, the testing circuit must also be constructed using such
devices. Among these devices, a Mach-Zehnder Interferometer (MZI)
stands out due to its high tolerance to temperature variations and
modulation capacity. MZIs can be designed either as 1$\times$1 input-output
(switching) devices using a Y-splitter and a Y-combiner, or as 
$2\times 2$ input-output {\it cross-bar switches} using waveguide
couplers and phase modulators. We propose an MZI-based architecture
for the testing circuit. Moreover, ours is a specification-based
testing approach, testing signal power and phase against a
reference. Since our DFT approach requires both a test and a
reference signal, we opt for a $2\times 2$ MZI-based
circuit. Moreover, as we use the MZI as a modulator device, we refer
to it as Mach-Zehnder modulator (MZM) in the paper. Our MZM is
specifically designed to access signals at test points, and
subsequently, a Y-combiner is used as a comparator for testing against
a reference. 

We demonstrate the application of our DFT circuit and test approach on
two distinct PIC designs: i) On an optical neural network (ONN) chip
design from \cite{Hochberg:ONN17}. Such designs are composed of MZIs,
phase modulators/attenuators, waveguide splitters, etc., in a {\it
  feed-forward} interconnection topology. We show how a MZI in the ONN
can be replaced by our DFT MZM and testing can be performed. ii) On an
optical logic circuit from \cite{condrat2011logic}. Such circuits also
comprise MZIs, splitters, combiners, etc., however, the circuit
topology incorporates signal feedback loops. We show how DFT insertion
and testing can be performed for such designs with feedback loops to
detect faulty behaviour. 

To exploit the full potential of such a DFT methodology, 
automatic tools and techniques for test-point selection as well as
ATPG (modeling signal power and phase) are also required. With the
availability of such tools, the DFT methodology may also enable
post-fabrication tuning of targeted devices. However, these test-point
selection and ATPG techniques are part of future work, and are beyond the
scope of this paper.  

{\it Paper Organization:} The following section reviews previous
work. Section III describes the proposed DFT architecture. Section IV
describes the design, implementation and validation of the circuit
components that comprise the DFT architecture. Section V describes the
application of our DFT approach to PIC testing. Finally, Section VI
concludes the paper.

\section{Previous Work}
Several approaches have been suggested by researchers to automate testing in PICs. A notable effort in this direction is demonstrated by \cite{electrically_eeasable} for conducting wafer-scale testing of photonic circuits. This study employed the concept of electrically erasable optical input and output ports utilizing micro-heaters for post-fabrication trimming of Si PICs or programmable PICs.\\
\indent On-chip power monitoring and calibration in the photonic integrated circuits (PICs) are also important aspect of testing a PIC. The paper \cite {PATD} shows the integration of the photon-assisted tunneling detectors with an interferometry based phase-interrogator structure for on-chip phase monitoring.\\
\indent A novel method of using probe card to electrically access and characterize the device, enabling the calibration and cloning of filter-based devices is presented in \cite{testing_sum}. The power coming from the through port of the device-under-test is sensed by a photodetector, and this information is delivered to a micro-processor (µP), that generates the signal to change the working point of the heaters (i.e., their driving voltage). \\
\indent Silicon photonics technology platform requires continuous process optimization and design verification. To fulfill this need, a test station for semi-automatic wafer-level characterization of silicon photonics devices has been proposed in \cite{test_station}. Several features such as fiber alignment, insertion loss etc. have been incorporated that are aimed at optimizing the accuracy and reproducibility of the measurement results.\\
\indent The paper \cite{LVS_} proposed a layout versus schematic (LVS) flow that addresses the particular need of curvilinear feature validation (curved path length and bend curvature extraction). The proposed LVS flow ensures more reliable photonic layout implementation.

While the impact of manufacturing
uncertainties on PIC operation is well-known
\cite{chrostowski2014impact}, there have been few attempts to develop
defect and failure models \cite{nikdast:JLT16} and analysis techniques
to estimate the impact of manufacturing variations on PIC's
performance \cite{banerjee:IEEE-DT-22}. In addition to testing for
fabrication defects, PICs are also required to undergo {\it
  calibration (or tuning)} \cite{calibration9} \cite{calibration2}
\cite{calibration7} so as to bring its performance within the
tolerance limits of the specification. Despite some advancements
\cite{calibration10} \cite{calibration8}, this process is manual and
tedious, lacks automation and incurs a high cost. 

The methodologies and techniques explored in the above works
predominantly focus on enhancing performance. However, there remains a
gap in the research and development of design-for-test architectures
to simplify testing, diagnosis and calibration of PICs, as well as
test pattern generation to automate the testing process. To bridge
this gap, we introduce a silicon-photonic DFT architecture, and
demonstrate its application to fault detection in a PIC. 

\section {The Proposed DFT Architecture }
 \begin{figure}
\centering
\includegraphics[scale=0.3]{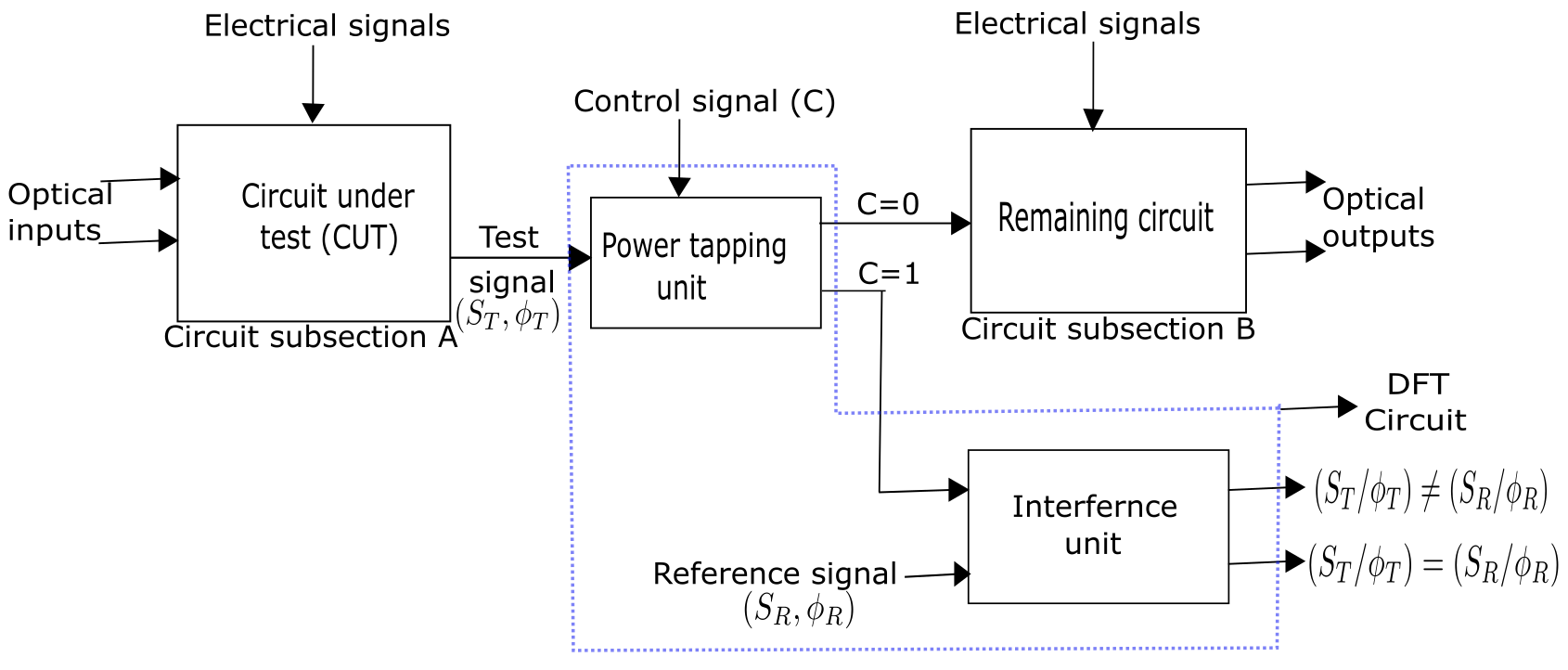}
\caption{Proposed DFT architecture}
\label{blockdiagram}
\end{figure} 


The proposed DFT insertion and test automation approach is shown in
Fig.\ref {blockdiagram}. A PIC with two circuit subsections $A$ and
$B$ is used to explain the proposed technique. Consider circuit $A$,
in the shown sample chip, as the optoelectronic circuit-under-test
(CUT) and its output  node, an optical debug point. The CUT has both
optical (data) and electrical (control) signals. The output signal of
circuit subsection $A$ with amplitude $S_T$ and phase $\phi_T$ is the
optical test signal. The test signal is fed into a power divider
circuit. The power divider circuit is operated by an (electrical)
control signal $C$. This circuit works under two modes 
such that under normal mode (say, $C=0$), the test signal is
transmitted to the circuit subsection $B$. Under the test mode (say,
$C=1$), a certain percentage of $S_T$ is fed into a DFT circuit. The
test signal is now tested against a reference signal of amplitude
$S_R$ and phase $\phi_R$. The DFT circuit compares the two signals,
and outputs whether $S_T$/$\phi_T$ is equal to $S_R$/$\phi_R$ or
not, where equality implies that the circuit is fault-free. It is
important to note that the DFT circuit is assumed to be either
lossless or its loss characteristics are known. For simplicity, we
also assume the proposed DFT circuit to be defect-free.   

\section{Design of the DFT circuitry}
The proposed DFT circuitry consists of a Mach-Zehnder modulator (MZM) for tapping power and a Y-combiner for comparing test signal with a reference signal. 
The MZM used in our architecture consists of directional couplers and a phase modulation stage.\\
 \begin{figure}
\centering
\includegraphics[scale=0.21]{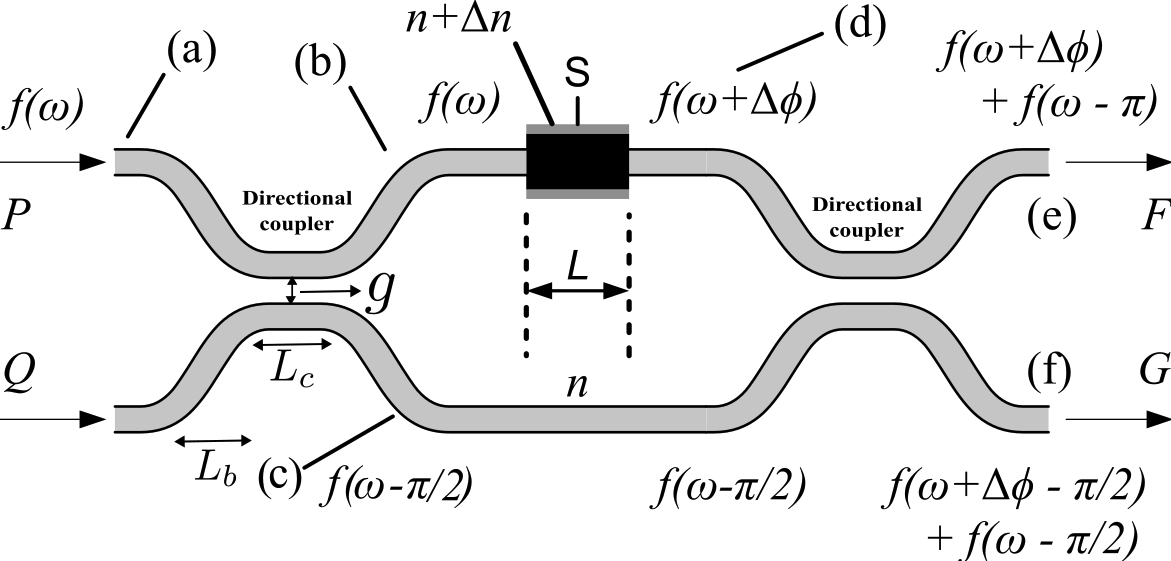}
\caption{Mach-Zehnder modulator}
\label{mzm}
\end{figure} 
\indent A MZM is a conventional integrated optic device used not only
for modulation, but also for routing through the use of coupling and
controlled interference. Consider the MZM depicted in Fig. \ref{mzm},
with optical inputs $P$ and $Q$, and outputs $F$ and $G$. The waveguides have
an index of refraction, $n$. Coupling arises when two waveguides are
brought into close proximity, allowing the electromagnetic fields of
one waveguide to extend over the other and vice versa. This
interaction results in energy transfer between the waveguides,
dependent on the coupling length. The couplers utilized in this device
are 3dB couplers, finely tuned to evenly distribute or combine the
signal from both inputs across the two outputs. In this configuration,
the signal at (a) traverses through the 3dB coupler, dividing between
outputs (b) and (c) due to coupling, inducing a phase shift of $\pi$/2
in waveguide (c) (note that the phase of the coupled signal lags the
original by $\pi/2$, as depicted).\\ 
\indent In the center region, S represents an external input employed
to modify the refractive index ($\Delta n$) of the upper
waveguide. This adjustment is achieved through various methods/devices
such as microheaters, carrier injection, advanced techniques like
high-speed MOS-capacitors, or alternative approaches. This change in
refractive index causes a path-length difference, and therefore a
phase difference, between the signals in (b) and (d). This phase
difference causes constructive or destructive interference at the
second coupler when the signals from (c) and (d) are combined. While
we have illustrated only one signal input at P, an input at Q
functions similarly. However, its output reaches the opposite output
compared to the input from P. The input applied at S may cause a phase
difference in the range of 0 to $\pi$, routing the input either
completely to one output, or distributing between two outputs or
completely to the other output, as depicted in Figs. \ref{fig:a} -
\ref{fig:c}. 

\begin{figure}[hbt]
\begin{minipage}[c]{0.3\linewidth}
\centering
\includegraphics[scale=0.23]{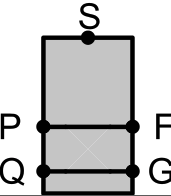}
\caption{Bar configuration}
\label{fig:a}
\end{minipage}\hfill
\begin{minipage}[c]{0.3\linewidth}
\centering
\includegraphics[scale=0.21]{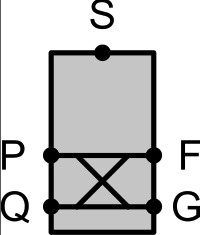}
\caption{Intermediate (test) configuration}
\label{fig:b}
\end{minipage}
\begin{minipage}[c]{0.3\linewidth}
\centering
\includegraphics[scale=0.21]{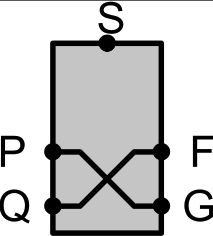}
\caption{Cross configuration}
\label{fig:c}
\end{minipage}
\end{figure}
\indent In our DFT MZM, we employ a carrier injection based phase
modulator at $S$ to effect modulation in the device. Specific bias
voltages at $S$ are applied to effect various MZM configurations. We
design the individual components (directional coupler, phase
modulator, and Y-combiner) and 
assemble them into a MZM-based DFT circuit. All the components are
designed and simulated using different 
suites of Ansys Lumerical \cite{Lumerical} such as the Finite
Difference Eigenmode (FDE) solver, Variational FDTD (varFDTD) solver,
Finite-Difference Time-Domain (FDTD), CHARGE, and INTERCONNECT. The
FDE solver calculates the mode field profiles, effective index, and
loss. The varFDTD solver is a 2.5D variational FDTD solver that
simulates the propagation of light in planar integrated optical
systems. The FDTD is a simulation software that helps design, model
and optimize photonic components. The Charge Transport (CHARGE) solver
is a physics-based electrical simulation tool for semiconductor
devices, which solves the system of equations describing the
electrostatic potential (Poisson’s equation) and density of free
carriers (the drift-diffusion equations). The INTERCONNECT simulates
classical and quantum PICs while enabling the co-design and
co-simulation of photonic and electronic circuits on multiple
electronic design automation (EDA) platforms \cite{Lumerical}. 

\subsection{Directional Coupler} 
\begin{figure}
\centering
\includegraphics[scale=0.3]{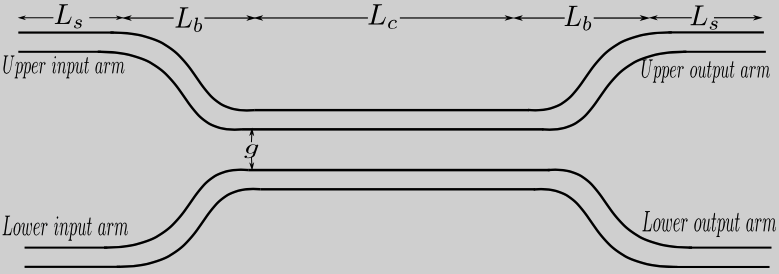}
\caption{Directional coupler schematic}
\label{fig1}
\end{figure}

We design a directional coupler as shown in Fig. \ref{fig1}. The input optical signal undergoes 100\% coupling from one waveguide to the other waveguide as it propagates forward. The optical signal is fed at the upper input arm of the  2X2 directional coupler. As the signal propagates in the device, and arrives at the upper waveguide of the coupling section, it starts to couple into the lower waveguide. The distribution of signal in the upper and lower arms at the outputs depends on the gap, $g$, between the waveguides in the coupling section and the length of coupling section, $L_c$. By choosing different values of $g$, and $L_c$, it is possible to achieve any desired fraction of the optical signal at the outputs of the waveguides. For our DFT, we aim at designing a 3dB directional coupler.\\
\begin{figure}
\centering
\includegraphics[scale=0.3]{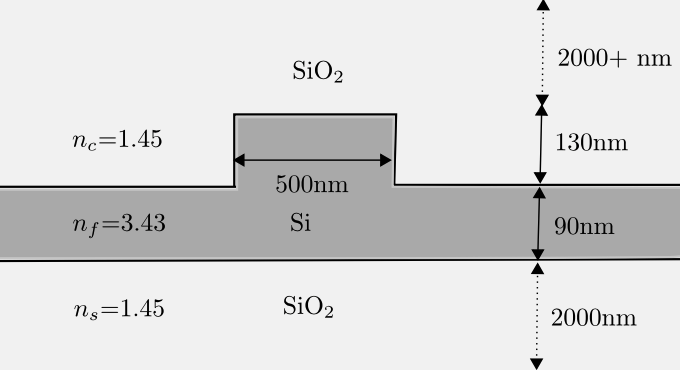}
\caption{Waveguide profile}
\label{wg}
\end{figure} 
\indent The waveguide profile used in our experiments is shown in Fig. \ref {wg}. The gap, $g$, is fixed at 200nm. The effective index of the initial mode, $n_o$ = 2.323, is calculated using finite difference eigen (FDE) mode solver. Upon introducing the second waveguide, we now have two modes, approximately represented by the sum and the difference of the original waveguides, and have effective refractive indices, $ n_1 = n_o + n/2$ and $n_2 = n_o - n/2$, where $n$ is the refractive index of silicon. We first calculate the coupling length by considering the difference in effective refractive indices, $\Delta n$, between the two coupled modes, such that $\Delta n$ = $n_1 - n_2 = 2.592793 - 2.544243 = 0.04855$. The coupling length, $L_c= 15.96 \mu m$, for a complete cross-over of optical signal is calculated as,
\begin{align}
 L_c = \frac{\lambda_o}{\pi \Delta n} sin ^{-1}\Big(\sqrt {\frac{P_{out}}{P_{in}}}\Big),
 \end{align}
where $\lambda_o = 1.55\mu m$ is the wavelength of light used,
$P_{out}$ and $P_{in}$ are the power at the output and input arms
respectively. The design and simulation of the directional coupler is
performed in FDTD. The corresponding $E$-field distribution in the
directional coupler is shown in Fig. \ref{fig2}. Here, we notice a
small fraction of signal being reflected in the lower input arm. To
avoid this loss of signal, waveguide bends are smoothed by increasing
the length from $L_s= 3\mu m$ to $L_s = 4\mu m$. The $E$-field
distribution in the improved design is shown in Fig. \ref{fig3}.\\ 
\begin{figure}
\centering
\includegraphics[scale=0.15]{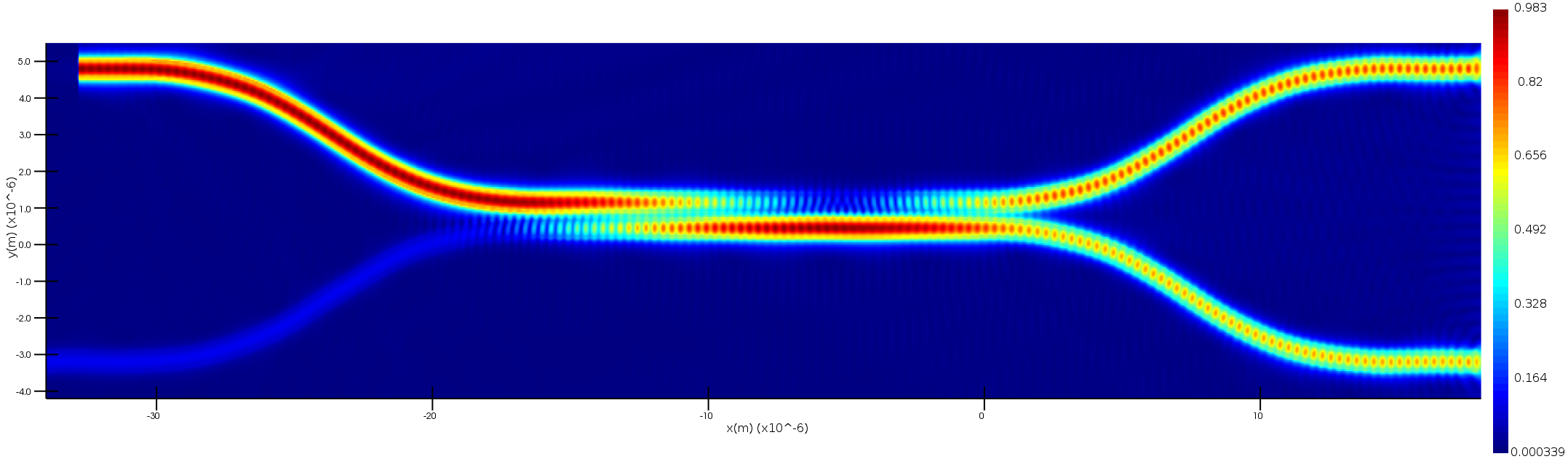}
\caption{Complete cross-over of E-field in the directional coupler. The color scheme represents E-field amplitude on a scale of 0 to 1.}
\label{fig2}
\end{figure} 
\begin{figure}
\centering
\includegraphics[scale=0.15]{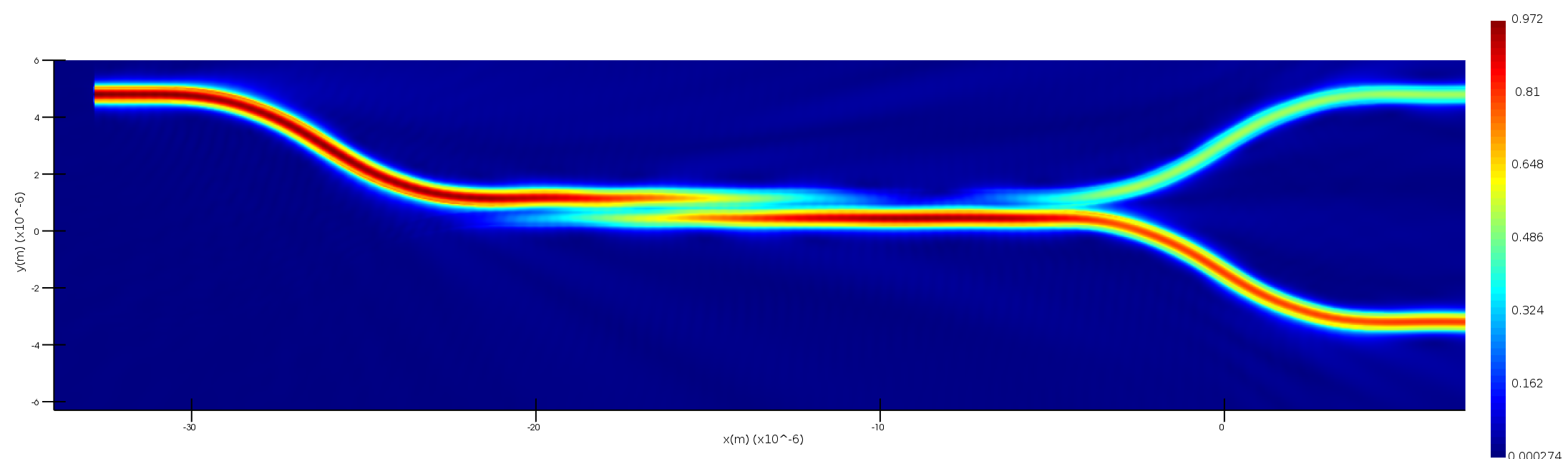}
\caption{Complete cross-over of E-field without reflection in a directional coupler}
\label{fig3}
\end{figure} 
\begin{figure}
\centering
\includegraphics[scale=0.62]{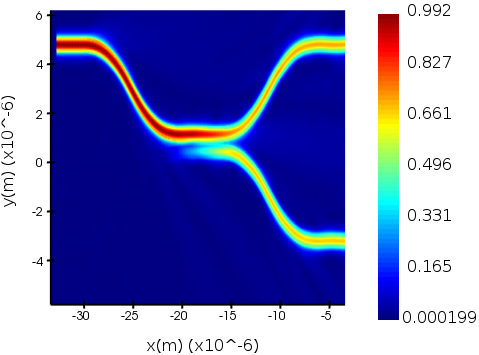}
\caption{E-field distribution in a 3dB directional coupler}
\label{fig4}
\end{figure} 
\indent As the coupling section ends and the waveguides move apart, the effective refractive index begins to change. As a result, a small fraction of the optical signal couples back into the upper output arm resulting in the change in signal distribution at the upper and lower output arms. The output power measured at the lower and the upper output arms are 89\% and 11\% respectively. Keeping this phenomenon in mind, a 50:50 power distribution will happen at a  different $L_c$ rather than a conventionally calculated $L_c/2$. We perform a sweep on  the length parameter, $L_c$, from 3$\mu m$ to 5$\mu m$. The result shows a 48:48 power distribution at 4.7$\mu m$  as shown in Fig. \ref{fig4}. The $s$-parameters of the 3dB directional coupler are extracted to be imported into INTERCONNECT at a later stage.

\subsection{Phase modulation}
\begin{figure}
\centering
\includegraphics[scale=0.42]{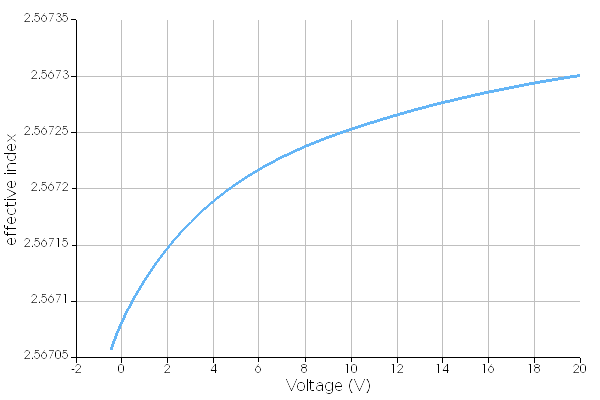}
\caption{Real part of $n_{eff}$ vs Voltage}
\label{fig5}
\end{figure}

\begin{figure}
\centering
\includegraphics[scale=0.42]{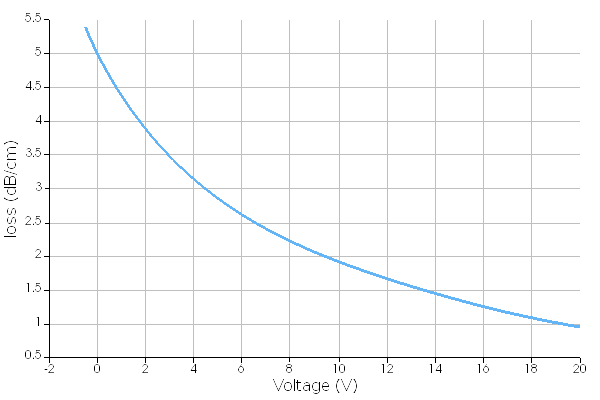}
\caption{Loss vs Voltage}
\label{fig6}
\end{figure}

\begin{figure}
\centering
\includegraphics[scale=0.42]{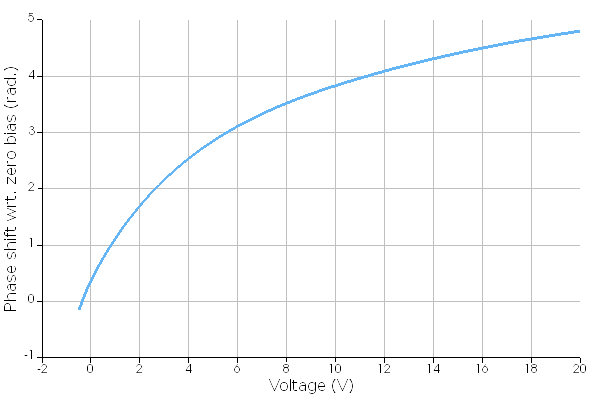}
\caption{Real part of $n_{eff}$ vs Voltage}
\label{fig7}
\end{figure}

The phase modulation characteristics of the MZM are obtained by
performing simulations in CHARGE and MODE solvers.  According to the
Drude expansion model, the refractive-index change, $\Delta n$, of
crystalline Si produced by an applied electric field, $E$, or by a
change in the concentration of charge carriers, $\Delta N$, is
calculated as, 
\begin{align}
\Delta n = \frac{ -(e^2\lambda_o ^2)}{8\pi ^2 c^2 \epsilon_o n} \Big[\frac{\Delta N_e}{m_{ce}^*}+ \frac {\Delta N_h}{m_{ch}^*}\Big],
\end{align}
 where $e$ is the electronic charge, $c$ is the speed of light, $\epsilon_o$ is the permittivity of free space, $n$ is the refractive index of unperturbed crystalline Si, $\Delta N_e$ is the change in concentration of electrons, $\Delta N_h$ is the change in concentration of holes, $m^*_{ce}$ is the conductivity effective mass of electrons, $m^*_{ch}$ is the conductivity effective mass of holes. \\
\indent The change in the concentrations of  free charge carriers affects the optical absorption coefficient of the material, $\Delta \alpha $, following the mathematical expression given as,
 \begin{align}
\Delta \alpha = \frac{ (e^3\lambda_o ^2)}{4\pi ^2 c^3 \epsilon_o n} \Big[\frac{\Delta N_e}{m_{ce}^{*2}\mu_e}+ \frac {\Delta N_h}{m_{ch}^{*2}\mu_h}\Big].
\end{align}
Here, $\mu_e$ is the electron mobility, and $\mu_h$ is the hole mobility \cite{optics:soref.carriers.electrooptic}. \\
\indent Experiments are performed to evaluate the effect of voltage on the optical signal traversing through silicon. A 5mm long Si waveguide is phase modulated by a reverse biased pn-junction driven by a 5mm long Al coplanar transmission line. The CHARGE solver provides charge density in the p-n junction with the changing reverse bias from 0 to 20V. The change in the charge density is imported into the MODE solver to calculate optical index modulation of the waveguide with the applied voltage. As the voltage is applied to the waveguide, $n_{eff}$ changes, resulting in the phase modulation of the optical signal. \\
\indent The results obtained are graphically represented in the effective refractive index (real part) vs voltage and loss vs voltage characteristics are shown in Figs. \ref{fig5} and Fig. \ref{fig6} respectively. The phase shift introduced in the optical signal with the applied voltage is shown in Fig. \ref{fig7}. The data obtained from these graphs are further imported to INTERCONNECT to be used for phase modulation in MZM.

\subsection{ Mach-Zehnder modulator (MZM)}
\begin{figure}
\centering
\includegraphics[scale=0.42]{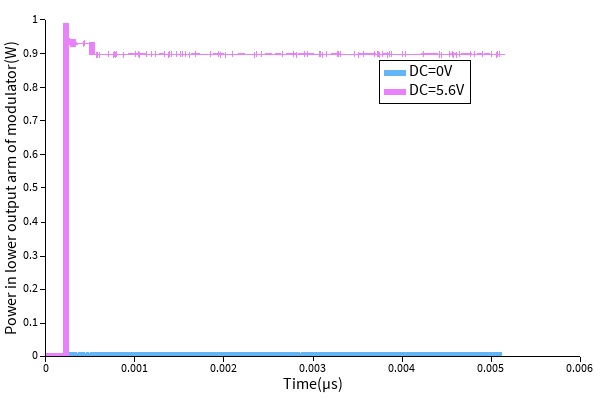}
\caption{MZM with 5mm modulator arm acting as switch with voltage application}
\label{fig8}
\end{figure}

\indent  A MZM is designed in the INTERCONNECT using the S-parameters of the 3dB coupler and $n_{eff}$  vs voltage data obtained in previous sub-sections respectively. The modulation section is 5mm long. The set up is powered by 1W continuous wave (CW) laser. At the output, the optical signals are measured using oscilloscopes. For 0V DC supply provided at the modulation stage,  the MZM works in a bar configuration. The complete coupling of optical signal to the lower output arm occurs with the application of 5.6V DC  supply at the modulation stage, as shown in Fig. \ref{fig8}, leaving MZM in a cross configuration. The optical signal at the output can be distributed between the two output arms using different voltage values in the range 0 to 5.6V.\\
\begin{figure}
\centering
\includegraphics[scale=0.45]{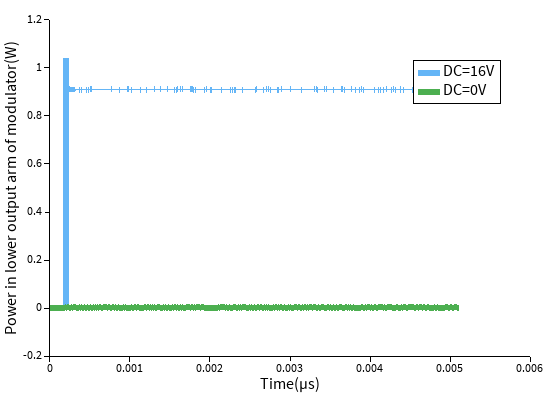}
\caption{MZM with 3mm modulator arm acting as switch with voltage application}
\label{3mzi}
\end{figure} 
\begin{figure}
\centering
\includegraphics[scale=0.42]{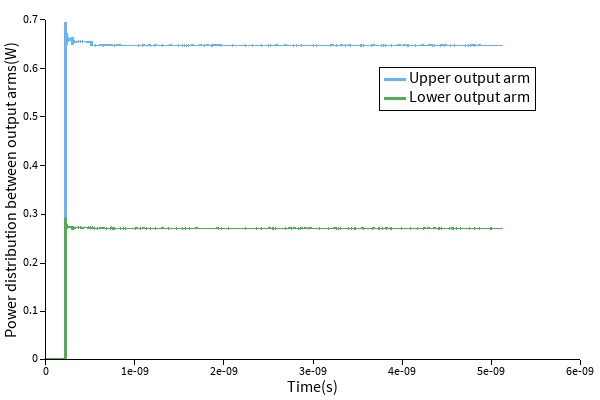}
\caption{Power distribution in MZM in intermediate configuration}
\label{figi}
\end{figure} 
\indent We designed another MZM with a shortened modulation arm length
of 3mm. It requires 16V to undergo switching from 0 to 1, as shown in
Fig. \ref{3mzi}. Keeping the footprint vs power tradeoff in mind, we
use the MZM with 5mm modulation arm length for our DFT architecture, as
the one with 3mm requires 16V for switching, which is impractical or
not readily available in conventional chip power supplies. \\ 
\indent The total optical signal at the outputs can be described as the sum of two waves,
\begin{align}
E(V_1,V_2)= \frac {E_o}{1+\sigma}\Big[\sigma e ^{\Big(\frac{-2\pi}{\lambda_o}n_{eff}(V_1)L\Big)} +  e ^{\Big(\frac{-2\pi}{\lambda_o}n_{eff}(V_2)L\Big)}\Big]
\end{align}
where $\sigma$ represents the splitting ratio (the ratio of the power in each branch), $L$ is the length of the arm, $\lambda_o$ is the free-space wavelength, and $n_{eff}(V)$ is the effective index of the arm as a function of the applied voltage. In the case where only one arm is actively driven,
\begin{align}
n_{eff}(V_2) = n_{eff,2},
\end{align}
and the phase accumulated in that arm is a constant,
\begin{align}
\phi_2 = \frac{2\pi}{\lambda_o}n_{eff,2}L
\end{align}
The normalized power transmission can then be calculated as
\begin{align}
T(V_1)&= \Big{| \frac{E(V_1)}{E_o}\Big|}^2{\nonumber}\\ 
&=\frac {1}{1+\sigma}{\Big|\sigma \exp (\frac{-2\pi}{\lambda_o}n_{eff}(V_1)L)+ \exp(-\sigma_2)\Big|}^2
\end{align}
The characteristics, $V_{\pi}L_{\pi}$ product, of MZM can be determined 
from,
\begin{align}
\Delta{\phi}= \frac{\Delta n_{eff}2\pi L}{\lambda_o}.
\end{align}
For $\Delta \phi = \pi$,
\begin{align}
\Delta n_{eff} (V_{\pi}) = \frac{\lambda_o}{2L_{\pi}}.
\end{align}
The insertion loss (IL) is defined as the ratio of the peak normalized transmission and the ideal transmission $T$=1. 
\begin{align}
IL (dB)= -10 log_{10}(\max T).
\end{align}
The extinction ratio (ER) is the ratio of the peak transmission power to the minimum transmission power, 
\begin{align}
ER (dB)= 10 log_{10}\Big(\frac{\max T}{\min T}\Big).
\end{align}
\\
The calculated values of IL and ER of the designed MZM are 2.6dB and 20.3dB respectively.\\
\subsubsection{The Test-Access Configuration of the MZM}
\indent We make use of the intermediate configuration of the MZM as
shown in Fig. \ref{fig:b}. In the intermediate configuration (test
mode), power is partially present in both arms. For a DC voltage of
1.77V, 30\% of the optical signal is coupled into the lower output arm
of the MZM, and 70\% in the upper arm, as shown in Fig. \ref{figi}. We
refer to this configuration as the {\it test mode} and the tapped-out
30\% signal as the test signal. The power and phase of this test
signal are measured using an oscilloscope.  
\subsection{Y-combiner}
\begin{figure}
\centering
\includegraphics[scale=0.52]{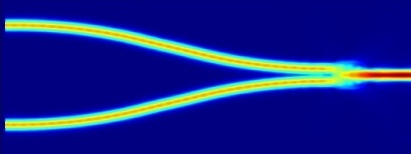}
\caption{E-field distribution in Y-combiner}
\label{fig9}
\end{figure}

\begin{figure}
\centering
\includegraphics[scale=0.36]{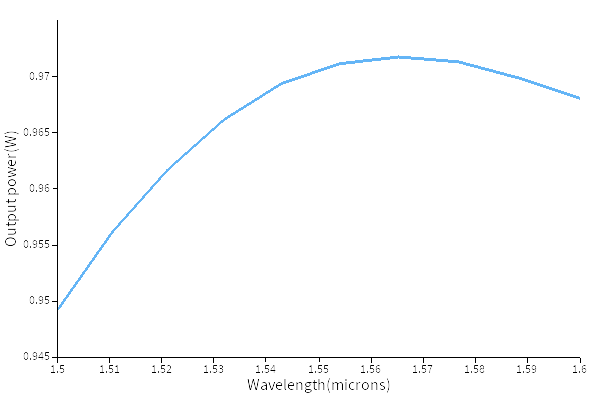}
\caption{Transmitted power at Y-combiner output}
\label{fig10}
\end{figure}

\indent A Y-combiner is designed as per the waveguide profile shown in Fig. \ref{wg}. The Y-combiner is 15$\mu m$ long and simulated in FDTD as shown in Fig. \ref{fig9}. The power transmission for 1.55$\mu m$ wavelength is shown in Fig. \ref{fig10}. The insertion loss calculated to be 0.22dB, is further used in INTERCONNECT for circuit simulation.\\
\indent The test signal obtained from MZM and a reference signal are
applied as the Y-combiner inputs. Here, the reference signal is the
ideal/fault-free optical signal at the test debug point. The power and phase of
the reference signal are known, or can be computed by analyzing the
circuit layout. The phase of the reference signal is equal and
opposite to that of the ideal (fault-free) test signal in a defect-free 
circuit. The output of the Y-combiner is measured using an
oscilloscope. If the output is zero, the CUT is defect-free, as it
implies that the test and reference signals canceled at the
Y-combiner due to destructive interference. Otherwise, the CUT is
defected. 

\begin{figure}
\centering
\includegraphics[scale=0.42]{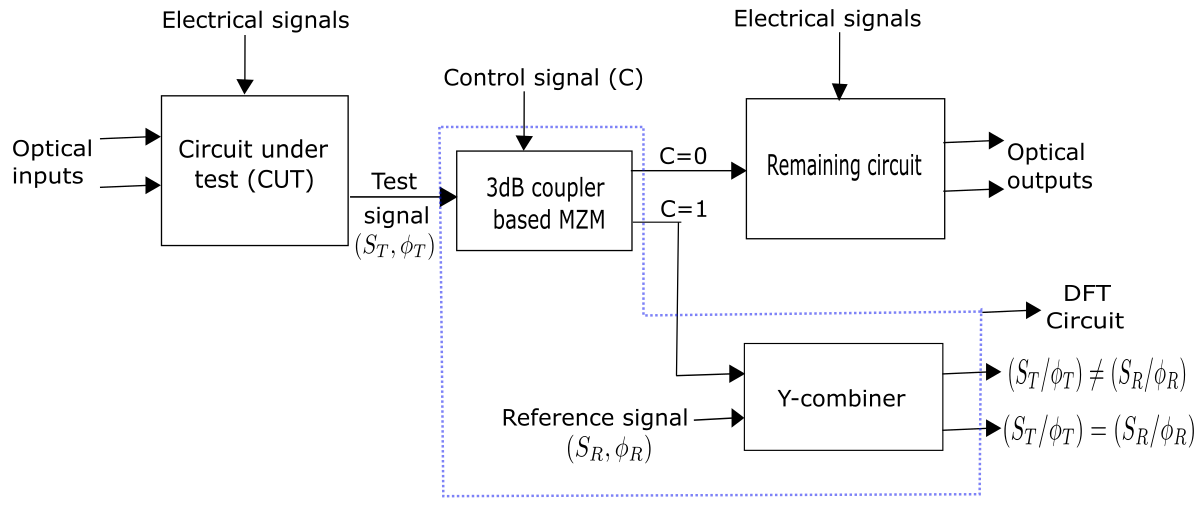}
\caption{Detailed block diagram of the DFT circuit}
\label{figf}
\end{figure}

A detailed block diagram of the DFT architecture composed of the
designed MZM and the Y-combiner is shown in Fig. \ref{figf}. A
tolerance-limit on the output of the Y-combiner, depending on CUT
applications, may be considered for decision-making.  

\section{Application of the DFT Approach}
The DFT unit serves as a critical sub-circuit facilitating comprehensive testing and debugging throughout the product lifecycle. Through meticulous design considerations, including strategic placement and functionality optimization, DFT techniques enhance testability, shorten time-to-market, and ultimately contribute to higher product quality and yield. The effective utilization of test pins and debug points can streamline the testing process and empower engineers to address complex design challenges with precision.\\
\indent The placement and operation of the proposed DFT circuit is
demonstrated in this section. We choose two widely used PIC
configurations and test them using our DFT circuit. 
\subsection{Testing a circuit with a feed-forward topology}
\begin{figure}[hbt]
\centering
\includegraphics[scale=0.67]{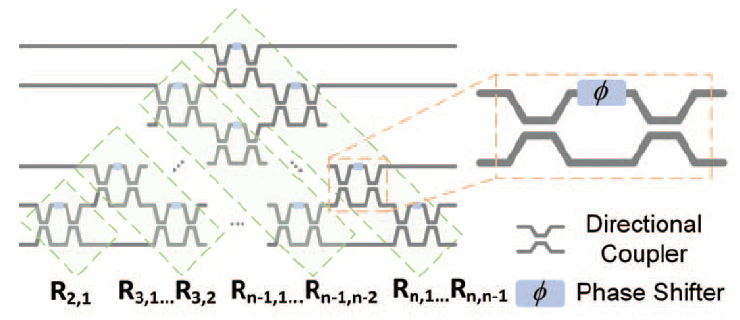}
\caption{Schematic of a triangular MZI array and the structure of a 2X2 MZI}
\label{neural_network}
\end{figure}

The classical integrated optical neural network (ONN) architecture
\cite{Hochberg:ONN17} implements MZI arrays to realize multi-layer
perceptron (MLP) inference as shown in Fig. \ref{neural_network}. The
plane rotator $R_{ij}$ is an $n\times n$ identity matrix, where four
entries at $(i, i)$, $(i, j)$, $(j, i)$ and $(j, j)$ indices are
replaced by $cos(\phi)$, $sin(\phi)$, $-sin(\phi)$, and
$cos(\phi)$. Each $R_{ij}$ can be implemented with a $2\times 2$ MZI,
whose transfer function is: 
  \begin{equation}
    \begin{pmatrix}
F\\ 
G
\end{pmatrix}
= 
\begin{pmatrix}
cos \phi & sin \phi \\
-sin \phi& cos \phi
\end{pmatrix}
\cdot
\begin{pmatrix}
P\\ 
Q
\label{eqn:1}
\end{pmatrix}
  \end{equation}
where the phase $\phi$ can be implemented with a optical phase
  shifter \cite{Hochberg:ONN17}. 

\begin{figure}[hbt]
\centering
\includegraphics[scale=0.67]{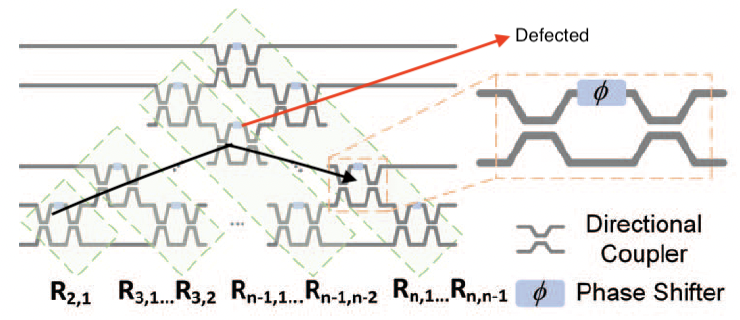}
\caption{Schematic of a triangular MZI array depicting optical signal propagation path and location of the deformed MZI}
\label{neural_network_copy}
\end{figure}
  
\indent This defect-free circuit topology is implemented in
INTERCONNECT. For a particular test configuration (test input), the
optical signal propagation path is shown in
Fig. \ref{neural_network_copy} with a black colored arrow. The MZI
marked with the orange dotted line is replaced by the proposed DFT MZM
circuit. Under normal mode of operation of DFT circuit, the optical signal passes
through the upper output (the $F$ arm) of the MZM. When  DC voltage of 1.77V is applied,
the DFT MZM operates under test mode and 30\% of the optical signal
(test signal) is coupled to the lower ($G$) arm. The power and phase
of the test signal is calculated to be 0.27 W and 18.7 rad using an
oscilloscope. A reference signal of 0.27W and phase -18.7rad, and also
the test signal, are applied as inputs to the Y-combiner. The two
signals destructively interfere and no signal is detected at the
Y-combiner output -- implying that the circuit is fault-free.

\begin{figure}[hbt]
\centering
\includegraphics[scale=0.45]{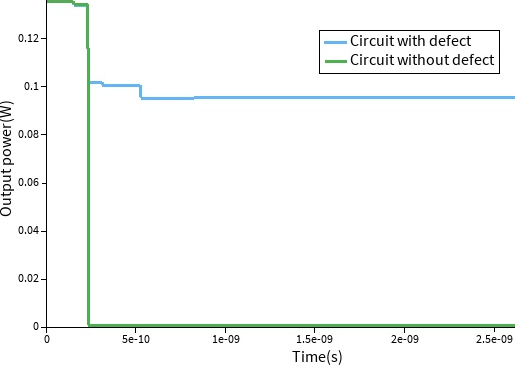}
\caption{DFT circuit output while testing neural network topology}
\label{nn}
\end{figure}

\indent Now, assume that a defect is introduced in the MZI marked with
a red arrow in Fig. \ref {neural_network_copy}. We may assume that the
manufacturing defect causes a phase shift of 2.14rad instead of
3.14rad in the MZM. The defect may result from physical deformities in
waveguides, material impurity or faulty modulation techniques. The
defect causes a change in the DFT MZM's output to 0.1W as shown in
Fig. \ref{nn} -- thus detecting malfunction (note that 0.1W is 
almost 30\% of the test signal's power). 

\subsection{Testing an optical logic circuit with feedback loops}

\begin{figure} [hbt]
\centering
\includegraphics[scale=0.3]{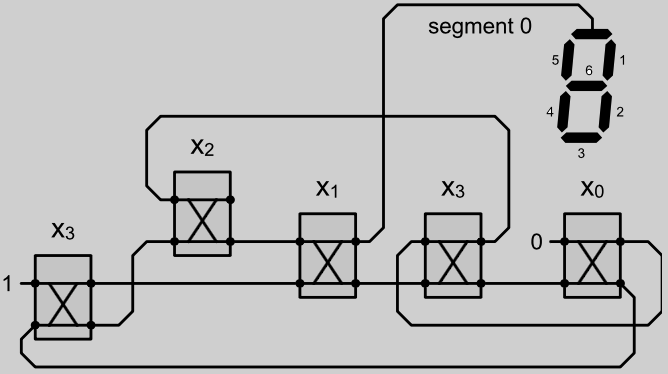}
\caption{Segment “0” of BCD-to-7 Segment Display}
\label{bcd7}
\end{figure}

\indent Fig. \ref{bcd7} implements a Boolean function corresponding to
Segment-0 of a BCD-to-7-segment display using 5 MZIs, which is
borrowed from \cite{condrat2011logic}. In this circuit, the MZIs act
as true cross-bar switches controlled by electrical signals, where the
control signal equal to 0 configures the MZI into a bar, whereas a 1
configures it into a cross. The optical signals ``1'' and ``0''
denoting laser inputs are connected to 
the MZI's data inputs. The shown interconnection composes the Boolean
function for Segment-0. The output is computed at the upper arm of the
MZI controlled by $X_1$, where a photodetector detects presence (1) or
absence (0) of light. Notice that under the input $X_3X_2X_1X_0$ =
0010, the waveguides of the circuit of Fig. \ref{bcd7} are connected
in a cycle/feedback loop. Thus, this optical logic design style
includes topological cycles.  

Consider the electrical control
input $X_3X_2X_1X_0$ = 0100 (corresponding to the decimal digit 4)
applied to the control the MZIs. Optical (laser) inputs 1 and 0 are
applied to the MZIs controlled by inputs $X_3$ and $X_0$,
respectively. This configuration should set
Segment-0 to 0.

\begin{figure} [hbt]
\centering
\includegraphics[scale=0.3]{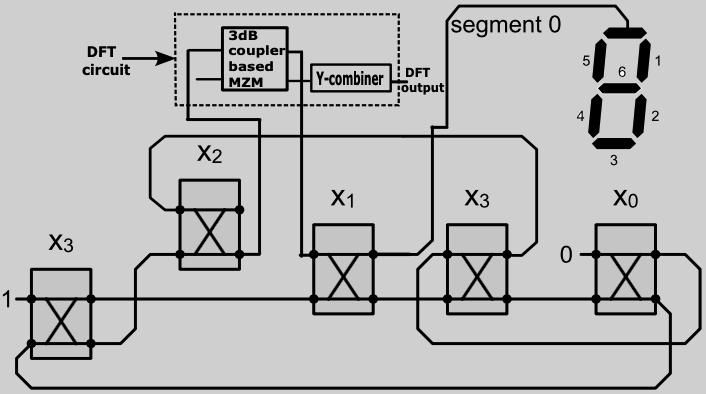}
\caption{DFT insertion in Segment “0” of BCD-to-7 Segment Display}
\label{bcd7_dft}
\end{figure}

\begin{figure}[hbt]
\centering
\includegraphics[scale=0.45]{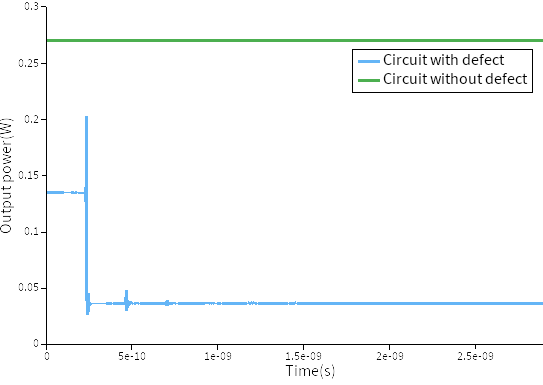}
\caption{DFT output for circuit with feedback loop}
\label{o-pfeedback}
 \end{figure}

We add the DFT circuitry between $X_2X_1$ as shown in
Fig. \ref{bcd7_dft}. In this case, the DFT circuit is inserted
externally in the BCD-to-7 segment display circuit because all four
ports of the constituent MZMs are used and replacing a MZM may disturb
the normal functioning of the 
given circuit. A reference signal of 0.27W and 0rad is applied to one
of the inputs of the Y-combiner. For a fault-free circuit, the
Y-combiner of DFT circuit should provide a 0.27W signal at the
output. In our experiment, we introduce a deformed MZI at the device
controlled by $X_3$ (the one on the right) by changing the phase shift
in the modulation arm from 3.14rad to 2.14 rad. The signal
distribution in the faulty MZI changes and a signal is detected at the
test debug point. Under test mode, 27\% of the detected signal moves
into Y-combiner and destructively interferes with the reference
signal. The output of the DFT circuit is shown in
Fig. \ref{o-pfeedback}, indicating the presence of a defect. 

\section{conclusion}
In conclusion, the emergence of Silicon (Si) photonics underscores the necessity for automated testing and validation techniques. Given the susceptibility of Si-photonics devices to manufacturing imperfections, the implementation of a Computer-Aided Design (CAD) based testing infrastructure becomes imperative to optimize testing workflows and support large-scale manufacturing. This study has introduced a methodology for testing photonic integrated circuits, proposing a design-for-test (DFT) circuit capable of comparing the test signal against a reference signal for the circuit under scrutiny. Through a series of experiments encompassing the design and simulation of all DFT architecture components, followed by circuit assembly, the effectiveness of the proposed DFT circuit has been demonstrated. Its utility has been explained through application on widely deployed photonic integrated circuits (PIC), where intentionally introduced defects were accurately evaluated using the proposed DFT circuitry.
\bibliography{test.bib}
\end{document}